\documentclass[useAMS,usenatbib]{mnras_tex_edited}

\usepackage{amsmath,array,amsfonts}
\usepackage{amssymb}
\usepackage{graphicx}
\usepackage{tabularx}
\usepackage{placeins}
\usepackage{float}
\usepackage{algorithm}
\usepackage[noend]{algpseudocode}
\usepackage{natbib}
\usepackage{pstricks}
\usepackage{verbatim}
\usepackage{subfigure}
\usepackage{booktabs}
\usepackage{siunitx,booktabs,multirow, dcolumn}
\usepackage{hyperref}
\usepackage{bm}

\newcommand{\ltsima} {$\; \buildrel < \over \sim \;$}
\newcommand{\gtsima} {$\; \buildrel > \over \sim \;$}
\newcommand{\lta} {\lower.5ex\hbox{\ltsima}}
\newcommand{\gta} {\lower.5ex\hbox{\gtsima}}

\begin{document}
\title{Binary radial velocity measurements with space-based
  gravitational-wave detectors}

\author[K.W.K.. Wong, V. Baibhav, E. Berti]
{\parbox{\textwidth}{Kaze W.K.~Wong$^1$, 
    Vishal Baibhav$^1$,
    Emanuele Berti$^1$
}\vspace{0.4cm}\\
\parbox{\textwidth}{$^1$Department of Physics and Astronomy, Johns Hopkins University, Baltimore, MD 21218 USA}}
\date{Accepted ;  Received ; in original form }
  
\maketitle

\begin{abstract}
  Unlike traditional electromagnetic measurements, gravitational-wave
  observations are not affected by crowding and extinction. For this
  reason, compact object binaries orbiting around a massive black hole
  can be used as probes of the inner environment of the black hole in
  regions inaccessible to traditional astronomical measurements. The
  orbit of the binary's barycenter around the massive black hole will
  cause a Doppler shift in the gravitational waveform which is in
  principle measurable by future space-based gravitational-wave
  interferometers, such as the Laser Interferometer Space Antenna
  (LISA). We investigate the conditions under which these Doppler
  shifts are observable by LISA. Our results imply that Doppler shift
  observations can be used to study the central region of globular
  clusters in the Milky Way, as well the central environment of
  extragalactic massive black holes.
\end{abstract}

\section{Introduction} 

The direct detection of gravitational waves (GWs) by the LIGO/Virgo
collaboration is the beginning of a new era in black hole (BH)
astronomy~\citep{LIGOScientific:2018jsj,LIGOScientific:2018mvr} and tests of
strong field gravity~\citep{2016PhRvL.116v1101A}.  All GW observations
so far constrained the properties of BHs (or neutron stars) in binary
systems. This is in stark contrast with ``traditional'' astronomical
BH observations, which rely on the interaction of isolated BHs with
the surrounding environment, such as nearby
stars~\citep{1995Natur.373..127M,Ghez:2008ms,Gillessen:2008qv} and
accreting
matter~\citep{Shakura:1972te,Peterson:2004nu,2010ApJ...725.1918O,Steeghs:2013ksa}.
By their very nature, these electromagnetic observations are subject
to modelling and systematic uncertainties, weakening the supporting
observational evidence for BHs and our ability to measure their
parameters. For example, the very existence of intermediate-mass BHs
(IMBHs) is still under
debate~\citep{2017Natur.542..203K,2017IJMPD..2630021M}.

Recent work considered various astrophysical processes, other than
cosmological redshift~\citep{Markovic:1993cr}, which may introduce
measurable Doppler shifts in gravitational
waveforms~\citep{Yunes:2010sm,Gerosa:2016vip,Meiron:2016ipr,Inayoshi:2017hgw,Randall:2018lnh},
and the astrophysical properties that could be inferred from such
measurements.  Doppler shift measurements in gravitational waveforms
extend the class of astrophysical systems that can be studied with GW
detectors beyond the strong-field merger and collapse of compact
objects.

Unlike electromagnetic measurements, GW measurements do not have
multiple emission lines that can be used to independently identify the
Doppler shift: an event moving at constant line-of-sight velocity is
degenerate with a heavier system without proper motion~\citep[see
e.g.][]{Flanagan:1997sx}.  For the proper motion to be detectable, we
need to observe {\em variations} in the line-of sight velocity.

One of the most common astrophysical systems that can produce
potentially detectable Doppler shifts are hierarchical triples. In
the hierarchical triple scenario, the orbital period of the binary
around the third body should not be too large compared to the the
observation period: if it is, the observed velocity of the binary will
be approximately constant during the observation, hence
indistinguishable from a system of different mass without proper
motion.  This also means that longer observation periods help us
resolve larger velocity variation timescales.  In contrast with
Earth-based interferometers, which typically observe binary inspirals
and mergers lasting for seconds or minutes, the Laser Interferometer
Space Antenna (LISA)~\citep{2017arXiv170200786A} will measure inspiral
events lasting as long as a few years, and it is therefore more
sensitive to Doppler shifts in the gravitational waveform.

With a few exceptions \citep{Randall:2018lnh}, most recent studies
considered LISA sources where the third body has mass comparable to
the GW-emitting
binary~\citep{Meiron:2016ipr,Bonvin:2016qxr,Robson:2018svj} or much
smaller than the GW-emitting
binary~\citep{Seto:2008di,Tamanini:2018cqb,Steffen:2018wpg}.
%
%
In this work we focus on the complementary scenario where the third
body is a BH of mass {\em much larger} than the GW-emitting binary,
and we ask: how close to the BH should the GW-emitting binary be in
order to yield meaningful constraints on the properties of the third
body?
We show that observationally interesting scenarios include (i) white dwarf
binaries (WDWDs) orbiting around IMBHs, (ii) stellar-origin BH
binaries (SOBHs) similar to those detected by LIGO/Virgo orbiting
around a nearby supermassive BH (SMBH) such as the one at our own
galactic center, and (iii) IMBH binaries orbiting around extragalactic
SMBHs.

The structure of the paper is as follows.  In section
\ref{sec:Analysis} we describe our model for the gravitational
waveform, and we review the Fisher matrix technique used in our
parameter estimation calculations.  In section \ref{sec:Result} we
present our main results. In section \ref{sec:Discussion} we discuss
the limitations of our work, their scientific implications, and
directions for future work.

\section{Doppler-shifted waveform model and parameter estimation} \label{sec:Analysis}

\begin{figure}
\includegraphics[width=\columnwidth]{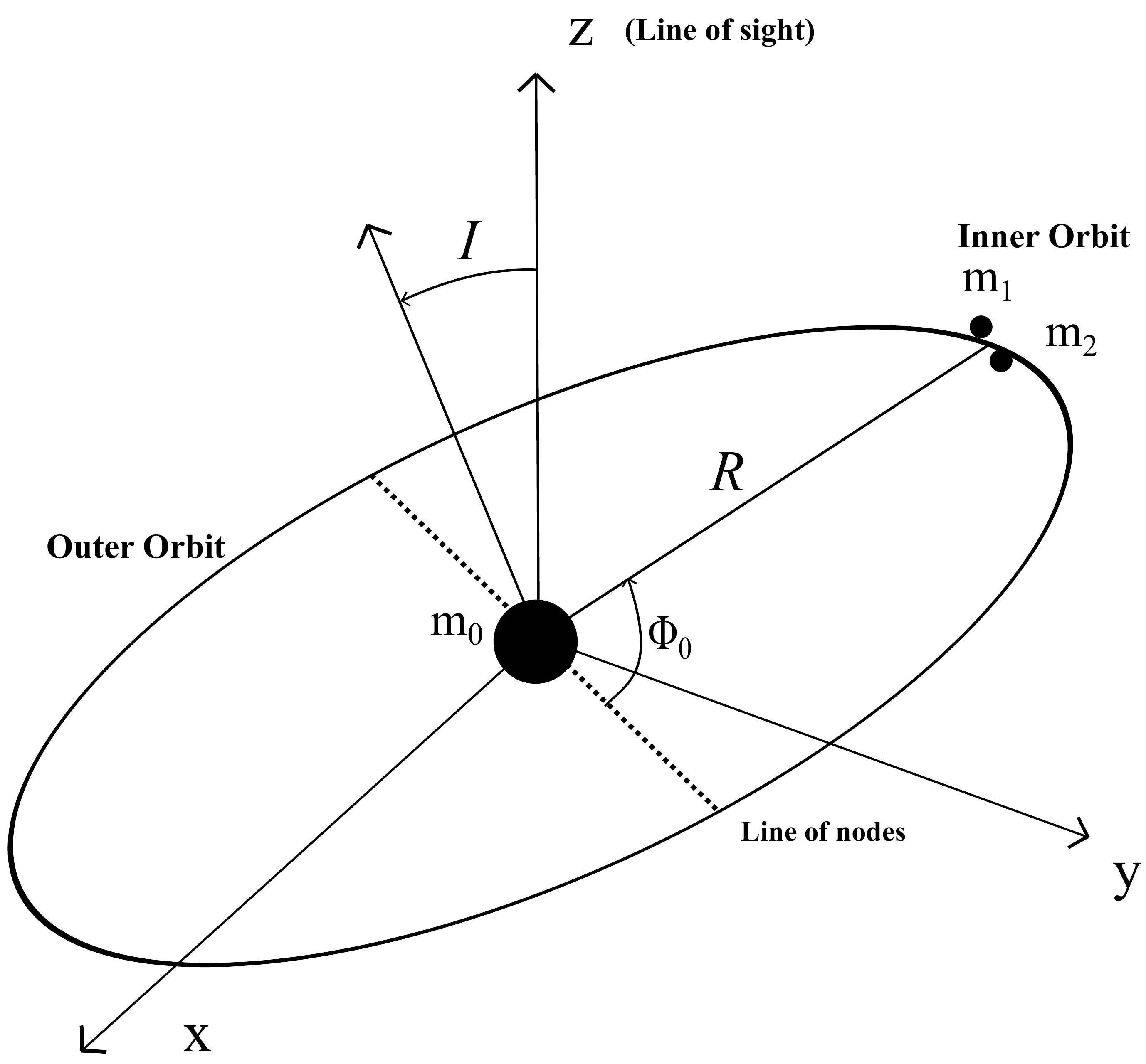}
\caption{Illustration of the geometry of the system.}
\label{fig:illustration}
\end{figure}

The geometry of the triple system we consider is sketched in
Fig.~\ref{fig:illustration}.  The $z$ axis is oriented along the line
of sight.  The GW-emitting binary components, with masses $m_1$ and
$m_2$, are on a circular ``inner'' orbit; $m_0$ is the mass of the
third body (a massive BH); $I$ is the angle between the line of sight
and the orbital angular momentum of the GW-emitting binary, whose
barycenter is assumed to be on a circular ``outer'' orbit about the
third body; $R$ is the radius of this circular orbit. Let
$M_{\rm tot} = m_0 + m_1 + m_2$ be the total mass of the triple.  We
assume that the separation between the components of the inner binary
is much smaller than $R$, i.e. that the period of the inner orbit is
much shorter than the period
\begin{align}
P_{0}= 2\pi \sqrt{\frac{R^3}{M_{\rm tot}}}
\label{eq:P0}
\end{align}
of the outer orbit.
Under this assumption, we can model the dynamics of the inner and
outer orbits separately.  The line-of-sight velocity $v(t)$ for a
distant observer located on the $z$ axis is
\begin{align}
v(t) = v_{||}\cos \left(\frac{2\pi t}{P_0}+\Phi_0\right),
\label{eq:velocity}
\end{align}
where
\begin{align}
v_{||} = \frac{m_0}{M_{\rm tot}}\frac{2\pi R\sin{I}}{P_0}
\label{eq:vpar}
\end{align}
is the magnitude of the line-of-sight velocity.  The initial observed
phase of the outer orbit $\Phi_0$ is equal to zero when the
GW-emitting binary is traveling along the line of sight.  Here and
below we use geometrical units ($G=c=1$).

Let us now consider the effect of the Doppler modulation on the
gravitational waveform.  We model the nonspinning, quasicircular
binary waveform by expanding the phasing up to second post-Newtonian
(2PN) order, including modulations due to LISA's orbital motion and
effects due to the source location in the
sky~\citep{Berti:2004bd}.  In the time domain, the waveform is
given by
\begin{align}
  h(t) &= \frac{2\mathcal{M}^{5/3}}{D_L}\left[\pi f(t)\right]^{2/3}
  \times\frac{\sqrt{3}}{2}\tilde{A}(t) 
  \cos \Psi_0(t)\,,
\end{align}
where
\begin{align}
\Psi_0(t) &=2\pi \int^t_0 f(t') dt' + \varphi_p(t)+\varphi_D(t)\,,
\label{eq:waveform_TD}
\end{align}
$f(t)$ is the GW frequency at time $t$ in the observer frame, and
$D_{L}$ is the luminosity distance of the source. Here $M = m_1 + m_2$
is the observed total mass, $\eta = m_{1}m_{2}/M^2$ the symmetric mass
ratio, and $\mathcal{M} = \eta^{3/5}M$ the chirp mass of the inner
binary. The factor $\frac{\sqrt{3}}{2}$ accounts for the fact that the
two independent LISA interferometer ``arms'' are at angles of
$60^{\circ}$.  The terms $\tilde{A}(t)$, $\varphi_{p}(t)$,
${\varphi}_{D}(t)$ are amplitude, polarization and Doppler-phase
modulations that arise from the orbital motion of LISA, respectively.  They can be
expressed as functions of the binary's orbital frequency $f$, sky
location $(\bar{\theta}_{S},\,\bar{\phi}_{S})$ and orbital angular
momentum direction $(\bar{\theta}_{L},\,\bar{\phi}_{L})$, where
overbars denote quantities in the Solar System barycenter frame.
Detailed expressions can be found in~\citep{Cutler:1997ta}.

The possibility to detect cosmological effects in gravitational
waveforms was discussed in great detail by \cite{Markovic:1993cr},
while the detectability of astrophysical Doppler shifts induced by
planetary systems around WDWD binaries was studied by
\cite{Seto:2008di}. The line-of-sight velocity changes the
observer-frame frequency of the source through a Doppler shift
$f_O = f_S(1+v)$. This results into an additional phasing term in the
waveform:
$\Psi_0(t) \to \Psi_0(t) + \phi_{\rm
  pm}(t)$~\citep{Seto:2008di,Bonvin:2016qxr,Inayoshi:2017hgw,Randall:2018lnh,Robson:2018svj},
where the proper-motion modulation $\phi_{\rm pm}(t)$ is related to
the velocity profile $v(t)$ of Eq.~\eqref{eq:velocity} by
\begin{align}
\phi_{\rm pm}(t) = 2\pi\int_0^t v(t')f(t') dt'\,.
\label{eq:phaseShift_accurate}
\end{align}
If we omit the effect of proper motion in the analysis of LISA data
this phase shift will appear as a residual, which contains information
on the properties of the outer orbit.

\begin{figure*}
\includegraphics[height=0.45\textheight]{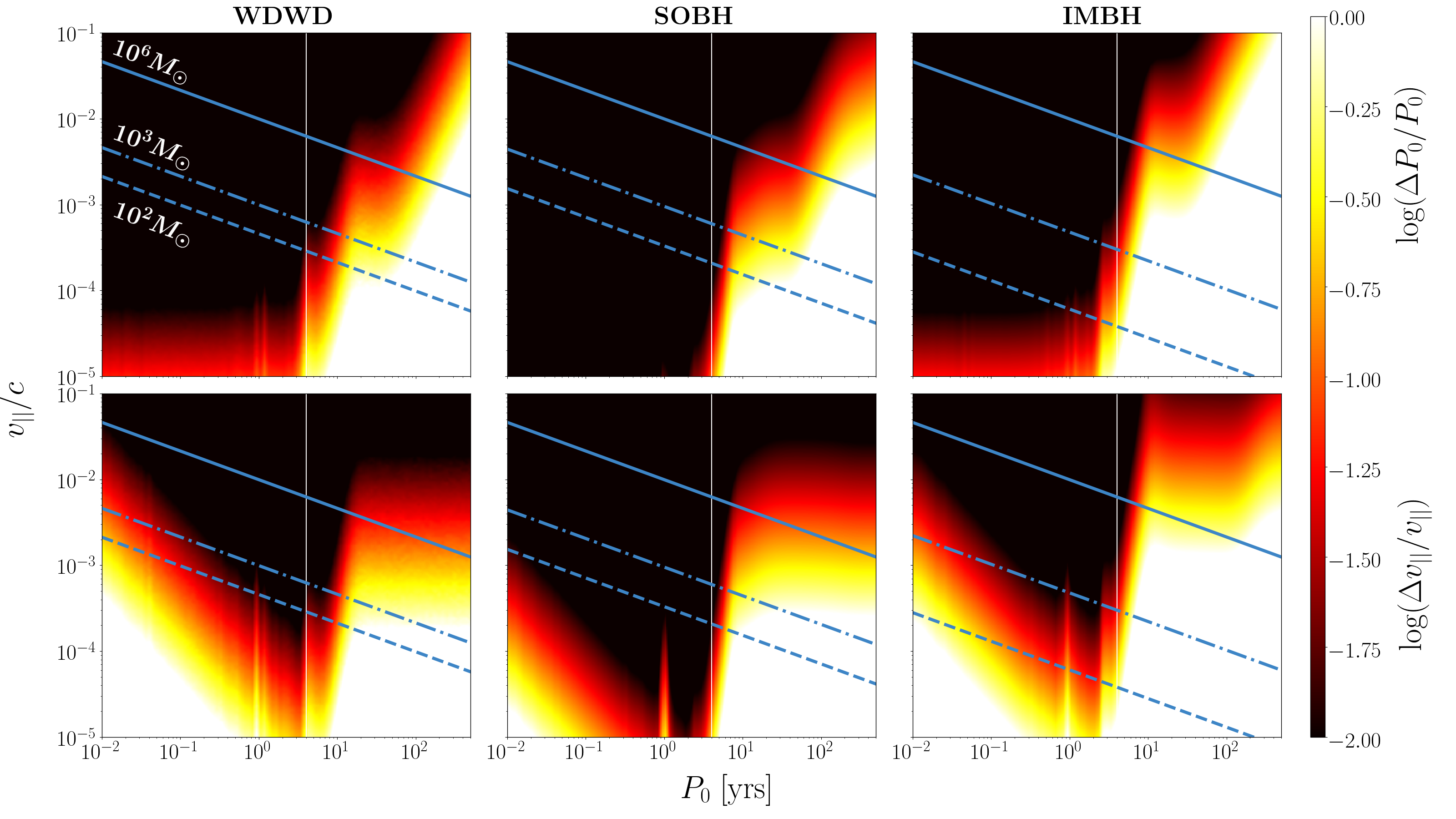}
\caption{Contour plots of the logarithmic relative error in the
  orbital period $P_0$ (top row) and magnitude of the line-of-sight
  velocity $v_{||}$ (bottom row) in the $(P_0,\,v_{||})$
  plane.  The three columns correspond to our three chosen physical
  systems at fixed SNR $\rho=10$: WDWD at $10^{-3} {\rm Hz}$ (left),
  SOBH (middle) and IMBH (right). In the regions above the solid,
  dash-dotted and dashed lines the third body must have mass larger
  than $10^6~M_{\odot}$, $10^3~M_{\odot}$ and $10^2~M_{\odot}$,
  respectively, to produce the observed Doppler shift. The vertical
  white line corresponds to the nominal LISA mission lifetime
  $T_{\rm obs}=4$~yr.  }
\label{fig:PeriodVelocityDelta}
\end{figure*}

Our waveform model depends on 12 parameters, which we will denote
collectively as ${\bm \theta}=\left\{\theta_i\right\}$. Nine of these
parameters -- i.e.
$\{D_L,\mathcal{M},\eta,t_c,\phi_c,\bar{\theta}_S,\bar{\theta}_L,\bar{\phi}_S,\bar{\phi}_L\}$,
where $t_c$ and $\phi_c$ are the coalescence time and
phase~\citep{Poisson:1995ef}, respectively -- characterize the inner binary; the
remaining three parameters $\{v_{||},P_0,\Phi_0\}$ characterize the
outer orbit.  Sampling the entire 12-dimensional parameter space is
computationally expensive. To estimate whether the outer orbit
parameters responsible for the Doppler modulation are measurable we use
a Fisher matrix analysis~\citep{Berti:2004bd}.  For a source with
signal-to-noise ratio (SNR) $\rho$, defined in terms of the LISA
one-sided spectral density $S_n(f)$ as
\begin{align}
\rho^2= 2 \sum_{\alpha=I,II} \int^{T_{\rm obs}}_{0}dt \frac{|h_\alpha(t)|^2}{S_n(f(t))}\,,
\end{align}
in the large-SNR approximation the uncertainties in the source
parameters are inversely proportional to $\rho$:
$\Delta \theta^i \propto 1/\rho$.  These uncertainties are given by
the diagonal terms of the covariance matrix $\Sigma$:
$\Delta \theta^{i} = \sqrt{\Sigma_{ii}}$
where $\Sigma=\Gamma^{-1}$ is the inverse of the Fisher information
matrix, with elements
\begin{align}
\Gamma_{ij}\equiv 2 \sum_{\alpha=I,II} \int^{T_{\rm obs}}_{0}dt
  \frac{\partial h_\alpha(t)}{\partial \theta^i}\frac{\partial
  h_\alpha(t)}{\partial \theta^j} \frac{1}{S_n(f(t))}\,.
\end{align}
Here $T_{\rm obs}$ is the observation time, and $\alpha$ labels the
two independent LISA data channels.

\section{Results} \label{sec:Result}

Sampling over the 12-dimensional parameter space is computationally
expensive, so we consider three specific examples to determine typical
conditions under which Doppler modulations may be detectable by LISA:
\begin{enumerate}
\item a $0.6\textrm{--}0.6~M_{\odot}$ WDWD binary with a source-frame GW
  frequency $10^{-3}~{\rm Hz}$;
\item a GW150914-like $36\textrm{--}29~M_{\odot}$ SOBH binary inspiral
  starting 5~yr before merger;
\item a $10^3\textrm{--}10^3~M_{\odot}$ IMBH binary inspiral starting
  5~yr before merger.
\end{enumerate}
All masses listed above are in the source frame.
We consider an observation time $T_{\rm obs}=4$~yr, corresponding to
the nominal LISA mission lifetime \citep{Audley:2017drz}.  We place
the sources at a luminosity distance such that the two-detector LISA
SNR is either $\rho=10$ or $\rho=100$. For convenience, these
luminosity distances are listed in Table~\ref{Tb:HorizonDistance}.
For simplicity we set $t_c=\phi_c=0$ and we choose the orientation
parameters to be
$\{\bar{\theta}_S,\bar{\theta}_L,\bar{\phi}_S,\bar{\phi}_L\} =
\{\arccos{(0.3)},\arccos{(-0.2)},5,4\}$ for all three sources.
We have verified that our choice of orientation parameters does not
significantly affect our conclusions (it mainly affects the
measurement accuracy by a rescaling of the SNR).

\begin{table}
  \caption{Luminosity distance for the three source classes considered
    in this paper.}
\begin{tabular}{llll}
\hline
  \hline  
$\rho$ & WDWD & SOBH & IMBH \\
\hline
10 & 1.40 kpc& 178 Mpc& 5720 Mpc\\
\hline
100 & 0.140 kpc & 16.9 Mpc& 679 Mpc\\
\hline
\end{tabular}
\label{Tb:HorizonDistance}
\end{table}

We sample over the outer binary parameters $v_{||}$ and $P_{0}$ within
the range $v_{||}/c\in [10^{-5},10^{-1}]$ and
$P_{0} \in [10^{-2},500]~{\rm yr}$. 
If $P_0<T_{\rm obs}$ the choice of the outer initial orbital phase
$\Phi_0$ does not significantly affect the measurement, since we can
measure a whole modulation cycle from the outer orbit, and therefore
we set $\Phi_0=0$.  If instead $P_0>T_{\rm obs}$ the choice of
$\Phi_0$ can affect the uncertainties and correlations between
parameters.  We postpone a more detailed investigation of this regime
to future work. For computational efficiency we compute parameter
estimation uncertainties following the frequency-domain method of
\cite{Chamberlain:2018snj} for inspiralling binaries (SOBH and IMBH),
while we use the time-domain procedure described in
Sec.~\ref{sec:Analysis} for WDWD binaries, where the inspiral is
negligible.

Figure~\ref{fig:PeriodVelocityDelta} shows uncertainties in $v_{||}$
and $P_0$ as a function of the simulated $v_{||}$ and $P_0$.  On the
left of each of the panels, $P_0<T_{\rm obs}$, and the correlation
between the parameters of interest is small.
In this case $f(t)$ is approximately constant, so we can pull it out
of the integral in Eq.~\eqref{eq:phaseShift_accurate} to find
\begin{align}
\phi_{\rm pm} = \frac{v_{||}P_0}{2\pi}\sin{\left(\frac{2\pi t}{P_0}+\Phi_0\right)}.
\end{align}
Then, ignoring correlations between parameters, the fractional
uncertainty on $v_{||}$ and $P_0$ scales as
\begin{subequations}
\begin{align}
\frac{\Delta v_{||}}{v_{||}} & \propto \sqrt{(\Gamma^{-1})_{v_{||}v_{||}}}\frac{1}{v_{||}} \propto \frac{1}{\rho v_{||}P_0}\,, \\
\frac{\Delta P_0}{P_0} & \propto \sqrt{(\Gamma^{-1})_{P_0P_0}}\frac{1}{P_0} \propto \frac{1}{\rho v_{||}}\,.
\end{align}
\end{subequations}
This is consistent with the behavior observed in
Figure~\ref{fig:PeriodVelocityDelta}.  The spike in the uncertainty on
$v_{||}$ occurs when $P_0\sim 1$~yr: in this case the Doppler
modulation due to the motion of the source is hard to measure because
it is degenerate with the Doppler phase from the motion of the LISA
detector~\citep[cf.][]{Tamanini:2018cqb}.

Just as in electromagnetic measurements based on the radial velocity
method, the observed velocity profile is completely degenerate with
inclination. Eq.~\eqref{eq:vpar} can be rewritten as
\begin{align}
\frac{m_0 R \sin{I}}{M_{\rm tot}} =\frac{P_0 v_{||}}{2\pi}\,.
\end{align}
Therefore $R$ and $\sin{I}$ (or $m_0$ and $\sin{I}$) cannot be
measured independently from Doppler shift measurements of $v_{||}$ and
$P_0$.  For any given measurement of $(v_{||},\,P_0)$ we can still
place a lower bound on $m_0$ (and a corresponding upper bound on $R$)
by setting $\sin{I} = 1$.  The blue lines in
Figure~\ref{fig:PeriodVelocityDelta} map the measured values of
$v_{||}$ and $P_0$ to the minimum mass of the third body
$m_0^{\rm min}$ necessary to produce the observed Doppler shift using
Eq.~\eqref{eq:vpar}.
For example, if we measure a signal consistent with
$v_{||} = 10^{-2} c$ and $P_0=2~{\rm years}$, the third body must have
mass $m_0> m_0^{\rm min}\sim 10^{3}~M_{\odot}$.

\begin{figure}
\includegraphics[width=\columnwidth]{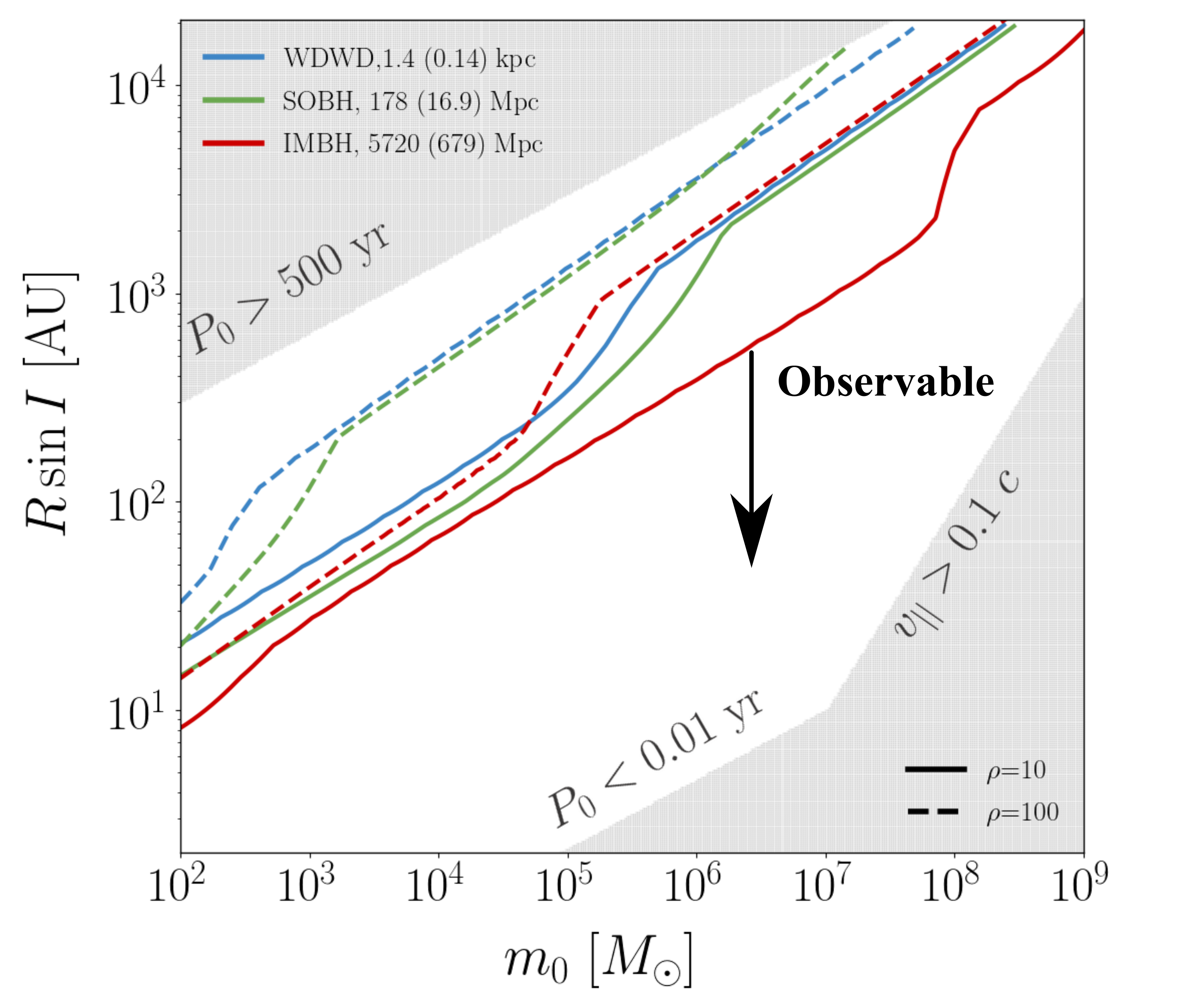}
\caption{Region in the $(m_0,\,R\sin{I})$ plane where the Doppler
  shift is observable: below each of the representative lines in this
  plot {\em both} $P_0$ and $v_{||}$ are measured to better than 10\%
  uncertainty.  The blue, green and red lines correspond to WDWD, SOBH and
  IMBH inner binaries, respectively. The solid lines refer to binaries
  that are barely detectable ($\rho=10$); the corresponding
  luminosity distances are given in Table~\ref{Tb:HorizonDistance} and
  in the legend. The dashed lines refer to loud detections with
  $\rho=100$.  The shaded regions were not sampled in our parameter
  estimation calculations (see the text).}
\label{fig:MoneyPlot}
\end{figure}

While in Figure~\ref{fig:PeriodVelocityDelta} we characterized our
ability to observe the binary's proper motion in terms of the
observables $v_{||}$ and $P_0$, from an astrophysical standpoint it is
more useful to use the mass of the third body $m_0$ and the outer
orbital radius $R$.  We can translate the results on the observability
of the Doppler shift in the $(v_{||},\,P_0)$ plane
(Figure~\ref{fig:PeriodVelocityDelta}) to criteria for the
observability of the Doppler shift in the
$(m_0,\,R \sin{I})$ plane, as shown in
Figure~\ref{fig:MoneyPlot}.
The shaded regions in this plot are not sampled in our parameter
estimation survey for one of the following reasons: the period $P_0$
is too long ($P_0>500$~yr, upper-left shaded triangle), too short
($P_0<0.01$~yr, bottom-right), or the magnitude of the line-of-sight
velocity $v_{||}>0.1$, so that the nonrelativistic approximation
becomes unreliable.

The solid and dashed lines show how close the inner binary can be to a
third body of mass $m _0$ for the proper motion Doppler signature to
be observable: below those lines, fractional uncertainties in {\em
  both} $v_{||}$ and $P_0$ are smaller than $10\%$ for binary signals
with SNR $\rho=10$ (solid lines) or $\rho=100$ (dashed lines).
The ``bumps'' in each of the solid and dashed lines correspond to the
slight plateau in the uncertainty on $P_0$ in the high-$P_0$ regime
(cf. the upper row of Figure~\ref{fig:PeriodVelocityDelta}).  For
astrophysical purposes, it is more useful to translate these SNR
values into the horizon distances listed in
Table~\ref{Tb:HorizonDistance} and to keep in mind that the source SNR
is inversely proportional to the luminosity distance (at least for
binaries in the local universe, such as WDWDs, where cosmological
effects are negligible). Recall also that for sources at distances where
cosmological effects are non-negligible (SOBHs and IMBHs) the observed
chirp mass of the source is redshifted to
$(1+z)\mathcal{M}$~\citep{Markovic:1993cr,Flanagan:1997sx}.

While we considered a wide range of possible values for $m_0$ and
$R\sin{I}$, it is clear from Table~\ref{Tb:HorizonDistance} that the
three source classes are of interest in different astrophysical
scenarios. WDWD systems can be detected by LISA within a few kpc, and
their Doppler modulation could be used to identify IMBHs in nearby
stellar clusters.  SOBHs can be detected within $\sim 200$~Mpc, and
their Doppler modulation could be used to probe the center of nearby
galaxies at $z\lesssim 0.05$. IMBHs can be detected out to
$\sim 6$~Gpc, so Doppler modulations could be used to study galaxy
formation out to redshifts $z\sim 1$. Note that the horizon distance
of the source has a strong dependence on the inner binary mass and sky
location (which were fixed in our study for computational reasons),
therefore the numbers quoted above should be considered as
illustrative for each astrophysical scenario, rather than as rigorous
detection limits.

\section{Discussion} \label{sec:Discussion}

This study is a proof-of-principle investigation of the conditions
under which Doppler shifts in gravitational waveforms could be
measurable by LISA.

Our analysis differs in several ways from recent work by
\cite{Inayoshi:2017hgw} and \cite{Bonvin:2016qxr}.
\cite{Inayoshi:2017hgw} carry out a Fisher matrix analysis using a
six-parameter model, including chirp mass, symmetric mass ratio,
distance, time and phase of coalescence and an acceleration parameter
$Y$. They do not account for LISA's orbital motion and the source
orientation, while we do. This is crucial: the measurability of
Doppler effects depends on the relative magnitude of the orbital
period of the source with respect to LISA's orbital period. There are
features (most notably the spike in uncertainties when the period of
the outer orbit is close to 1~yr) that can only be accounted for when
LISA's motion is considered. Besides, ignoring degeneracies
(e.g. between the inclination of the source and the chirp mass) can
lead to overly optimistic parameter estimation. Another important
difference is that the acceleration parameter $Y$ used in
\cite{Inayoshi:2017hgw} is a linear approximation of the phase shift
we considered here, so their phase shift is linear in time: see
e.g. their Eq.~(7). This corresponds to a special case of our
analysis: the long-period regime. In this regime the correlation
between the velocity of the source and the period of the outer orbit
is important, but it was ignored in \cite{Inayoshi:2017hgw} . Besides,
a measurement of their parameter $Y$ cannot be translated into a
measurement of the third body's mass and of the orbital radius of the
binary $R$ around the third body: at least one more variable is needed
to get a lower bound on the mass of the third body.

\cite{Bonvin:2016qxr} and \cite{Inayoshi:2017hgw} consider mostly
SOBHs as gravitational-wave sources, while we also considered WDWD
binaries and IMBH binaries. As we show in our work, the detectability
of Doppler effects depends dramatically on the choice of
source. Besides, both \cite{Inayoshi:2017hgw} and
\cite{Bonvin:2016qxr} focus on the acceleration effect. The values of
the $\epsilon$ parameter introduced in Eq.~(51) of
\cite{Bonvin:2016qxr} correspond to orbital periods $\gtrsim 10^4$~yr,
well beyond the range considered in our study.

In this exploratory study we have made simplifying
assumptions that we discuss below, and that should be relaxed in more
realistic scenarios.

The assumption of a circular outer orbit can be considered
conservative, because eccentricity in the outer orbit makes Doppler
shifts easier to observe \citep{Robson:2018svj}.  In general, there
will be a trade-off between the detectability gain due to large
variations in $v_{||}$ near pericenter and the fact that (because of
Kepler's second law) it is statistically more likely to find
astrophysical systems near apocenter.  The distribution of the outer
and inner orbital eccentricities plays an important role when
computing rates of LISA events with observable Doppler shifts
\citep{Nishizawa:2016jji,Breivik:2016ddj,Randall:2017jop,Randall:2018nud,Randall:2018lnh,Samsing:2018isx,DOrazio:2018jnv,Rodriguez:2018pss,Nishizawa:2016eza},
and we plan to address this issue in future work.

We modelled the outer orbit using Newtonian dynamics. This should be
sufficient for most astrophysical systems of interest: the dominant
corrections to the equations of motion enter at order $(v/c)^2$, and
therefore they should be mostly negligible even for
$v_{||}\sim 0.1 c$.  Furthermore, the dominant post-Newtonian
correction increases the orbital period \citep[see e.g.][]{2014grav.book.....P},
hence it improves the observability bounds shown in
Figure~\ref{fig:MoneyPlot} for given orbital parameters. In
this sense, once again, our predictions are conservative.

In principle we can convolve the Doppler observability criteria shown
in Figure~\ref{fig:MoneyPlot} with astrophysical models to predict the
number of events for which LISA will be able to observe Doppler
shifts. Vice versa, we could use LISA observations of Doppler shifts
(or the lack thereof) to constrain astrophysical models.  A detailed
discussion of the astrophysical implications of our results is beyond
the scope of this paper. In the hope to stimulate further research, we
briefly discuss some astrophysical scenarios that could lead to
observable Doppler shifts for the three source classes considered in
this paper: WDWD, SOBH and IMBH binaries.

\noindent {\bf (i) WDWDs:} WDWD systems can be detected by LISA within
a few kpc, and their Doppler modulation could be used to identify
IMBHs in nearby stellar clusters.  There is a broad range of estimates
of the number of binaries detectable by LISA in Milky Way globular
clusters, with some of the latest estimates ranging from a few to tens
of events \citep{Kremer:2018tzm}. The uncertainties are dominated by
assumptions on cluster models, such as the binary
fraction~\citep{Ivanova:2005mi,Sollima:2007sc,Hurley:2007pv,Albrow:2001ud}
and the efficiency of different dynamical
processes~\citep{Henon1971,AmaroSeoane:2001hv,Fregeau:2004if}.  The
number of WDWD events with an observable proper motion signature in
LISA could be used to set constraints on these cluster models.

\noindent {\bf (ii) SOBHs:} SOBHs can be detected by LISA within
$\sim 200$~Mpc, and their Doppler modulation could be used to probe
the center of nearby galaxies at $z\lesssim 0.05$.  The LIGO/Virgo
collaboration has already detected 10 BH-BH binary
mergers~\citep{LIGOScientific:2018jsj}, and yet there is no consensus
on the astrophysical origin of these
mergers~\citep{LIGOScientific:2018mvr}.  One possibility is that these
compact binaries are formed in the vicinity of
AGNs~\citep[see e.g.][]{2018arXiv181110627F}.  Indeed, the presence of X-ray
binaries~\citep{Hailey2018} and hypervelocity
stars~\citep{Brown:2005ta,Sherwin:2007tz} close to our galactic center
indicates that a large number of binaries exist in galactic nuclei.
Gaseous drags or three-body interactions in AGN disks can lead to very
hard binaries that should coalesce within a Hubble
time~\citep{Stone:2016wzz}.  \cite{Hoang:2017fvh} showed that the
merger rates of such binaries could be comparable to other dynamical
channels.  The merger process is very efficient if these binaries lie
within $\sim 0.1$~pc, resulting in a significant fraction of mergers
happening very close to the SMBH.  Furthermore, compact objects
embedded in AGN disks create density perturbations, resulting in
torques that lead to inward migration of the compact object.
Sometimes this torque changes sign, leading to the formation of
migration traps at $40 M$-- $600 M$ from the central
objects~\citep{Bellovary:2015ifg} which could act as hotbeds for the
formation of BH-BH binaries.  In summary, there are various scenarios
that could lead to SOBH binaries merging very close to an SMBH. These
systems may have detectable Doppler shifts that could serve as smoking
guns for binary formation in AGNs. It is even possible that SMBH mass
measurements from Doppler-shifted GWs could complement and/or improve
electromagnetic estimates of the AGN mass.

Another interesting scenario was proposed by \cite{Han:2018hby} and
\cite{Chen:2018axp}. These authors proposed that some extreme
mass-ratio inspirals (EMRIs) could actually be {\em binary} BH systems
inspiralling into a supermassive BHs. These ``binary EMRIs'' (or
b-EMRIs) source GWs both through the motion of the inner BH-BH binary
and through the inspiral of the b-EMRI. If such sources exist, the
Doppler shift in the GWs from the BH-BH binary could allow us to
estimate the SMBH mass. The Doppler-shift estimate of the central SMBH
mass could be used as an independent check of the parameters estimated
using the gravitational radiation from the b-EMRI inspiral.

\noindent
{\bf (iii) IMBHs:} IMBH binaries can be detected by LISA out to a few Gpc,
so Doppler modulations could be used to study galaxy formation out to
redshifts $z\sim 1$.
Despite claims of a connection between IMBHs and ultra-luminous X-ray
sources~\citep{Colbert:1999es} and other observational
evidence~\citep{Caballero-Garcia:2013vxa,Pasham:2015tca,2018ApJ...863....1C,2019arXiv190105496N},
there is still no conclusive observational confirmation of the
existence of IMBHs \citep[see e.g.][for a
review]{2017IJMPD..2630021M}.  IMBH detections could bridge the gap
between SOBHs and SMBHs, and help us understand how SMBH were born and
grew.  In some scenarios, clusters containing IMBHs sink towards the
galactic nucleus through dynamical friction, and upon evaporation
deposit their IMBHs near the galactic center
\citep{Ebisuzaki:2001qm}. The IMBHs then form binaries and eventually
merge, forming an SMBH. Some of these IMBH binaries could end up in
orbit around a more massive central
object~\citep{2018ApJ...856...92F,2018ApJ...867..119F}, and orbital
Doppler shifts could lead to biases in their estimated
masses~\citep{2018MNRAS.477.4423A,2019MNRAS.483..152A}. GW detections
of these systems by Earth- or space-based interferometers could
provide conclusive evidence of SMBH formation through runaway IMBH
collisions.

In conclusion, several astrophysical GW sources are expected to form
triple systems where the main GW emission from an ``inner'' orbit is
affected by Doppler modulations due to the ``outer'' orbital motion of
the binary around a third body.  LISA (unlike ground based detectors)
may observe the radiation from the inner binary for months or
years. GW searches and parameter estimation methods rely on waveform
modelling, so failure to account for Doppler modulations could
introduce systematics in parameter estimation and reduce the
efficiency of GW searches \citep{Bonvin:2016qxr}.  In this paper we
argued that, more interestingly, these effects may be observable,
enabling GW detectors to probe weak-field astrophysical processes
through the Doppler modulations of their strong-field inspiral
dynamics.  We investigated the conditions under which LISA may place
meaningful constraints on the third body's properties, and we
identified and discussed some classes of astrophysical systems of
particular interest as observational targets.

We plan to extend the present research in two main directions: (i) by
developing more realistic (and complex) possibilities for the orbital
motion and GW emission of the triples, and (ii) by using astrophysical
models to identify the most promising astrophysical systems that could
lead to LISA detections of Doppler modulations.

\section*{acknowledgments}
We thank Nicola Tamanini, Davide Gerosa, Hsiang-Chih Hwang, Tyson Littenberg,
Christopher Moore, Lisa Randall, Travis Robson and Nadia Zakamska for
useful discussions.  K.W.K.W., V.B. and E.B. are supported by NSF
Grant No. PHY-1841464, NSF Grant No. AST-1841358, NSF-XSEDE Grant
No. PHY-090003, and NASA ATP Grant No. 17-ATP17-0225.  This research
project received funding from the European Union's Horizon 2020
research and innovation programme under the Marie Skłodowska-Curie
grant agreement No. 690904, and it used computational resources at the
Maryland Advanced Research Computing Center (MARCC).  The authors
would like to acknowledge networking support by the GWverse COST
Action CA16104, ``Black holes, gravitational waves and fundamental
physics.''

\bibliographystyle{mnras_tex}
\bibliography{ProperMotion}

\end{document}